\documentclass[prb,twocolumn,showpacs]{revtex4}\usepackage{graphicx,color}\newcommand{\N}{\noindent}
\begin{document}
\title{Model of fragmentation of the exciton inner ring in semiconductor quantum wells}
\author{A.A.~Chernyuk}\altaffiliation{Institute for Nuclear Research, National Academy of Sciences of Ukraine (47, Nauki Ave., Kyiv 03680, Ukraine)}\email{inr@ukr.net}\author{V.I.~Sugakov}\author{V.V.~Tomylko}\affiliation{Institute for Nuclear Research, NAS of Ukraine, Kyiv, Ukraine}
\begin{abstract}The appearance of the non-homogeneous structures of the indirect exciton density distribution in the region of the quantum well (in the region of the inner ring) is explained. The structure (the fragmentation) occurs due to the exciton condensed phase formation because of interaction between excitons. The formation of the structure is related with the non-equalibrity of the system, which is caused by the exciton finite lifetime and the presence of the pumpimg. The structure emerges in the shape of a set of islands or circles of the condensed phase. The structure type depends on the pumping intensity, the size of the laser spot and disappears with increasing the temperature. The merging of two structures, created by different laser spots, is investigated at decreasing the distance between the centers of the spots.\end{abstract}
\maketitle
\section{Introduction}This work was fullfiled in connection with the experimental observation of the structure appearance in the exciton density distribution in double quantum well in the region of the laser spot\cite{1}.

Experimental studies of indirect excitons in GaAs/AlGaAs heterostructures, carried out over last ten years, found unexpected non-trivial effects. Under the influence of a strong electric field created by the external electrodes perpendicular to the wells, an electron and a hole, which form so-called "indirect exciton", are spatially separated (they are located in different wells). The lifetime of such an exciton is long, thus it is possible to create high concentrations of indirect excitons and to observe collective phenomena. One of non-trivial phenomena found in the exciton luminescence spectra from the double quantum wells, irradiated by the laser, is the emergence of inhomogeneous structures (sometimes periodical) in the exciton density distribution. For example, a periodic distribution of islands of the emission was found\cite{2} along the ring, which was arranged around the laser spot at a great distance (several hundred microns) from the center of the laser spot. At the study of the emission from a quantum well, excited by the laser through the round window in the electrode\cite{3}, the islands in the emission were observed, too, and they were periodically arranged along a ring under the perimeter of the electrode. Many experimental and theoretical studies of various aspects of the emerging spatial structures have been published
\cite{1,2,3,5,6,Par07,8,Lev05,Cher06,10,11,12,13,14,Sug06,Sug07,17,Sap08,Cher09,21,Muk10,23,24,25,Wil12,And12,Sug11,Sug14,Cher12,Bab13,Ivan}
at present. The different structures were observed in the configurations of the aperture in the electrode in the shape of a square, of a rectangle\cite{25}, a pair of related circles\cite{25}, a periodical potential distribution\cite{21} in the plane of the quantum wells.

Manyfold  theoretical models were proposed to explain the appearance of the structures of the exciton density distribution \cite{8,Lev05,Cher06,Par07,Sug06,Sug07,17,Sap08,Cher09,Muk10,And12,Sug11,Sug14,Wil12,Cher12,Bab13}. The main efforts were applied for the ascertainment of the principal possibility of the appearance of the periodicity of the exciton density distribution without a detailed application for the experiment. Specific explanation of the experiment was given in the works \cite{Lev05,Wil12} with respect to only one experiment \cite{2} and in the series of our works \cite{8,Cher06,Sug06,Sug07,17,Cher09,Cher12} applied to different experiments. The authors of the work \cite{Lev05} considered the instability, which arises under kinetics of  level occupations by the particles with the Bose-Einstein statistics. Namely, the growth of the occupation of the level with zero moment should stimulate the transitions of excitons to this level. But the density of excitons was found greater, and the temperature was found lower than these values observed on the experiments. In the
paper \cite{Wil12} the authors did not take into account the screening between charges on macroscopic distances.

In our model we consider the structure formation as a self-organization process at the phase transitions in the system of unstable particles. The appearance of a periodical distribution
of the exciton density, due to the finite lifetime and an attractive interaction between exccitons, was predicted in Ref.~\cite{Sug86}. The model of the phase transitions of particles with a finite lifetime successfully managed to describe \cite{8,Cher06,Sug07,17,Cher09,Sug11,Sug14,Cher12} the emission from the
ring around the laser spot, observed in Ref.~\cite{2}, the emission from the quantum well under the aperture in the electrode, observed in Refs.~\cite{3,23,25}, the structure appearance in the exciton system arising in Ref.~\cite{21}.

The main assumptions of the model are the following.

\N{\bf1)} There is an attraction between the particles that causes a condensed phase formation.  The interaction between indirect excitons was investigated by the quantum mechanical calculations in  several papers \cite{Fil,30,31,32,33}. The interaction consists of the long-range dipole-dipole repulsion and also of the exchange and Van-der-Waals short-range attraction. The dipole-dipole interaction grows with increasing the distance $d$ between the quantum wells due to the increase of the exciton dipole moment. At small distances between wells, the attractive interactions exceed the dipole-dipole repulsion and the exciton-exciton interaction energy is negative in a certain range of the distances between excitons or it experiences a dip in the dependence of the interaction energy on the distance between excitons (see Fig.~5 in Ref.~\cite{31} and Fig.~6 in Ref.~\cite{33}). According to Ref.~\cite{31}, the dip disappears at $d>$1.1$a_{\rm X}$, where $a_{\rm X}$ is the exciton radius. If the condition $d<$1.1$a_{\rm X}$ is obeyed, a biexciton may exist in the system at small values of $d$, when the total energy of the exciton-exciton interaction is negative, or the probability for two excitons to be a certain distance apart has a maximum due to the presence of the dip in interaction. We suggest that, in this case,  a condensed phase may arise in a multi-exciton system. The existence of a new phase in indirect exciton system was obtained in Ref.~\cite{Loz}, where the multi-exciton system was considered by the quantum field methods taking into account the dipoile-dipole, exchange and Van-der-Waals interactions for the case $T$=0. It was shown in Ref.~\cite{Loz}, that the exciton liquid phase may exist at $d<$1.1$a_{\rm X}$ in a stable phase and till $d<$1.9$a_{\rm X}$ in a metastable state. However, the assumption of an attractive interaction between the indirect excitons was criticized in the experimental work \cite{Yan07}, which showed that the frequency position of the emission maximum of the island of the condensed phase, where the density of excitons has maximum, is shifted to shorter wavelengths in the comparison with the frequency of the emission from the region between the islands, where the exciton density is less. The offset is small, it is about several times smaller than the emission bandwidth. Nevertheless, its presence may indicate a repulsion between excitons, and this conclusion is stated in Ref.~\cite{Yan07}. This result comes into conflict with our suggestion about the attractive interaction at some distances between wells. This contradiction was removed  in the papers~\cite{Sug11,Sug14}, where it was shown that the result of the work\cite{Yan07} may be explained by the presence of saturating localized states in the quantum well. The explanation is the following. At low pumping of the radiation, the emission is formed by low-lying localized states. With growth of the pumping, localized states become saturated, and after that the states of free excitons begin to be populated. When the exciton density becomes high, islands of the condensed phase are formed. The emission from the region between the islands corresponds to the emission from the localized states, the emission of the islands corresponds to the emission from the condensed excitons, whose levels are situated higher than levels of localized excitons. The lowering the energy due to the condensation is small, it is less than the total bandwidth formed by the both free and localized excitons. The free excitons can not move to the localized states, because  these states are occupied. This explains the frequency distribution in the spectrum observed in Ref.~\cite{Yan07}, in which the emission from condensed phase has higher frequency in comparison with the emission frequency from the region, where the condensed phase is absent.

\N{\bf2)} A finite lifetime of excitons plays am important role in the formation of spatial structures of the exciton density. Typically, the time of establishing of a local equilibrium is much less than the lifetime of an indirect exciton, and  a  local equilibrium state is established very fast, in which the properties of the exciton system in the neighborhood of some point in the space depends on the exciton density at a given point. However, in the presence of the phase separation, which occurs in the studied system there is a new time parameter, namely the time of establishment of the equilibrium distribution between the
different phases. In fact, it is time of equilibrium establishment in the entire macroscopic system. This time is large, it is associated with the diffusion motion of excitons over macroscopic distances. In practice, the equilibrium is not established, and the formed structures depend on the exciton lifetime (i.e., the regions of new phases grow as much as the exciton lifetime allows). If the exciton lifetime is so small, that during its lifetime an exciton can not be displaced by a distance of the order of the radius of the new phase nucleus, a new phase may not be formed \cite{34}. Namely, the lifetime determines the size, position and shape of islands. The considered structures are non-equilibrium, due to the finite lifetime and the pumping presence. Such structures, that exist in non-equilibrium conditions,  are called dissipative according to Ref.~\cite{35}. Thus, in our approach, a structure with high density of excitons is a consequence of the formation of the condensed phases in non-equilibrium conditions, i.e. the formation of the structures are the processes of self-organization in non-equilibrium systems. At the same time, the condensed phase arises as a result of the attraction between excitons and it is not due to their condensation into a state (with $\mathop k \limits ^\to=0$, for example), i.e. not due to Bose-Einstein condensation. Bose statistics of excitons is not necessary to describe the structure, although it may be important for determining the parameters characterizing the condensed phase.

During the last  several years, publications have appeared, in which coherence was observed in the emission from different regions of an island of the condensed phase or even from different islands \cite{10,11,Yang,Fogler,High}.  It was shown in Ref.~\cite{High} that the coherence length in the regions of the macroscopically ordered exciton state on the external ring  is much larger than in a classical gas, indicating a coherent state with a much narrower distribution  in the momentum space (than for the classical excitons). Our model is phenomenological, it describes the distribution of the exciton density, and  therefore cannot explain the interference effects in wave function. Yet, the model is successful in explaining the appearance of various exciton structures observed experimentally as well as their properties.

In Ref.~\cite{2} it was shown that a so-called inner ring exists, besides an outer one. The inner ring is observed as the emission from the region in the vicinity of the laser spot. Explanation of the origin of the inner ring was represented in the work \cite{Ivan} taking into account the relaxation of the exciton subsystem. The confirmation of the model proposed in Ref.~\cite{Ivan} was given in Ref.~\cite{1}. It was shown in this work that an irradiation of the inner ring created by some laser by another weak laser can lead to intensity emission reduction, despite of the increase of the exciton density. The effect occurs due to local heating, which induces a reduction of optically active excitons. Also in Ref.~\cite{1}, a fragmentation of the inner ring was found. However, the explanation of the fragmentation of the inner ring remains unknown.

In this paper, using the model of non-equilibrium phase transition, which is described above, we show the existence of fragmentation of the rings emission observed in Ref.~\cite{1} in the laser spot region, and consider the transition from a continuous ring to a fragmented one. Also the merging of two structures is described in case of irradiation by two lasers while their spot centers converge.

\section{Model of the system. Basic equations}We shall calculate the distribution of the exciton density in the quantum well plane under stationary irradiation by a Gaussian-shaped spot of the intensity distribution and on assumption that a gas phase and a condensed one can exist in the exciton system. In the previous studies of the condensed phase, two popular models of the phase transitions were used. We applied the stochastic model of the nucleation-growth (Livshits-Slyosov) in the works\cite{8,Sug06,17}  and the model of spinodal decomposition (Kahn-Hillert) in the works \cite{Cher06,Sug07,Cher09,Sug11,Sug14,Cher12}, modified and generalized for a system of particles with a finite lifetime. For the solution of the  problem of the present paper, we applied a modified model of spinodal decomposition.

The conservation law for the exciton density $n$, living a finite time $\tau$, is
\begin{equation}\label{eq1}\frac{\partial n}{\partial t}=-{\rm div}\mathop j\limits^ \to+G-\frac{n}{\tau }\end{equation}

\N where $G$ is the pumping (the number of excitons produced per time unit per square unit in the quantum well plane), $\mathop j \limits^ \to=M \mathop \nabla \limits^ \to\mu$ denotes the
exciton density, $\mu$ is the chemical potential, $M=nD / \left(k_{\rm B} T\right)$, $D$ is the diffusion coefficient. In the case of local equilibrium, the system can be described by the free energy, which depends on the exciton density. The chemical potential can be expressed in terms of the free energy $\mu=\delta F / \delta n$. The free energy is chosen in the Landau model:
\begin{equation}\label{eq2}F\left[ n \right] = \int {d\mathop r \limits^ \to\left[ {\frac{K}{2}\left( {\mathop {\nabla n} \limits^ \to} \right)^2 + f\left( n \right)} \right]} .\end{equation}
The member $\frac{K}{2}\left({\mathop {\nabla n}\limits^ \to} \right)^2$ characterizes the energy of heterogeneity. The free energy density $f$ is approximated as
\begin{equation}\label{eq3}f\left( n \right)=f_0+kTn\left({\ln \frac{n}{n_0}-1}\right)+\frac{an^2}{2}+\frac{bn^3}{3}+\frac{cn^4}{4},\end{equation}

\N where the term $kTn \left ({\ln n / n_0 - 1 } \right)$ prevaluates at low densities of excitons, describing their diffusive motion far from the laser spot, and the members of a power series expansion for the exciton concentration play a major role for large values $n$, meanly they describe the emerging
structure. The parameters $a$, $b$, $c$ in equation (\ref {eq3}) are chosen so that the free energy as a function of the density has a minimum corresponding to the condensed phase and describes the spectral shift to higher frequencies with increasing $n$ due to the dipole-dipole repulsion. To accomplish this, the following conditions should be valid: $a>0$, $b<0$, $c>0$. The value $a\cdot n$  is of order of the concentration shift of the spectra due to the dipole-dipole interaction. The chemical potential defined by the free energy (\ref{eq3}) enlarges with increasing the density. At some range of the density, the chemical potential decreases that corresponds to the creation of a condensed phase. Further, with increasing the pumping, it shifts to the violet side. The decrease of the energy determines  the energy gain at condensation per exciton. In the experiment~Ref.\cite{3}  the narrow emission line of indirect excitons, which arose when the laser intensity exceeded a certain threshold and may be interpreted as the emission from the condensed phase, experiences firstly a red shift with increasing the pumping . The maximum value of the red shift reached 0.5 meV. This value is less than the bandwidth of indirect excitons, observed in Ref.~\cite{2} (the value of order of meV). Similarly to the chemical potential, obtained from the free energy (\ref{eq3}), the narrow line shifts to the violet side with further increasing the pumping~\cite{3}. So, in the double quantum wells system based on AlGaAs the energy gain at condensation per exciton is small, much smaller than the gain at creation of the electron-hole liquid per electron-hole pair in silicon and germanium (of order of several tens meV \cite{Rice}), where new emission line is observed at condensation.

In dimensionless variables with units
\begin{equation}\label{eq4}l_{\rm u} = \sqrt {\frac{K}{a}} ,\quad n_{\rm u} = \sqrt {\frac{a}{c}}, \quad t_{\rm u} = \frac{k_B TK}{Da^2n_{\rm u}}, \quad G_{\rm u}=\frac{n_{\rm u}}{t_{\rm u}} \end{equation}
\N the equation for the exciton density (\ref {eq1}) can be rewritten:
\begin{eqnarray}\label{eq5}\frac{\partial n}{\partial t} = d_1 \Delta n + \mathop \nabla \limits^ \to \left[ {n\mathop \nabla \limits^ \to \left( { - \Delta n + n + b_1 n^2 + n^3} \right)} \right]\nonumber\\+G - \frac{n}{\tau }.\end{eqnarray}

\N where $b_1$=$b/\sqrt{ac}$.

Let believe the pumping spot to have as a Gaussian form:
\begin{equation}\label{eq6}G = G_0 \exp \left( { - \frac{\rho ^2}{2\sigma ^2}} \right),\end{equation}
\N where $\sigma$ is the parameter describing the width of the laser spot, $G_0$ is the value of the pumping at the maximum of the spot. Total number of produced particles pumped per time unit
equals to $2\,\pi\,\sigma^2G_0$.

Equation (\ref{eq5}) is a nonlinear 2D phenomenological equation describing the distribution of the exciton density of high concentration in the quantum well plane, taking into account the inhomogeneous pumping, the finite lifetime of the particles, the diffusive motion of particles and assuming the presence of a
condensed phase. Next, we shall give numerical results of its stationary solutions depending on the pumping. Numerical calculation methodology has been tested in previous works. In particular, the calculations were performed on a square lattice, whose dimensions were chosen such that the resulting solution does not depend on them. The independence of solutions on increasing the size of the computational domain, on the boundary conditions and unchanging the solutions over large times were checked, too. The initial density was chosen equal  zero. The simulation was performed by a Runge-Kutta method.

\section{Calculations of the exciton density with increasing pumping for different widths of the laser spot}
Let analyze the distribution of the exciton density as a function of the pumping for different radii of the laser spot. Firstly, let regard a narrow beam, the radius of which  is compared to the size of an island of the exciton condensed phase (according to experimental data, the island radius of the condensed phase is the value of order of few microns). The dependence of the exciton density distribution on the pumping (\ref{eq6}), which was obtained by solving the equation (\ref{eq5}), is presented in Fig.~1. At low pumping, the exciton density exists only in the region under the laser spot and corresponds to the gas phase  (Fig.~1a). With increasing the pumping, an region of high exciton density arises by a stick-slip way (Fig.~1b), and it believed to be an island of the exciton condensed phase.  With further increasing the pumping, this region expands, remaining the height of the exciton density almost unchanged (Fig.~1c-d). This indicates the presence of a condensed phase. So, in the case of narrow beams the condensed phase appears, but a fragmented structure does not occur.

\begin{figure}\centerline{\includegraphics[width=8.6cm]{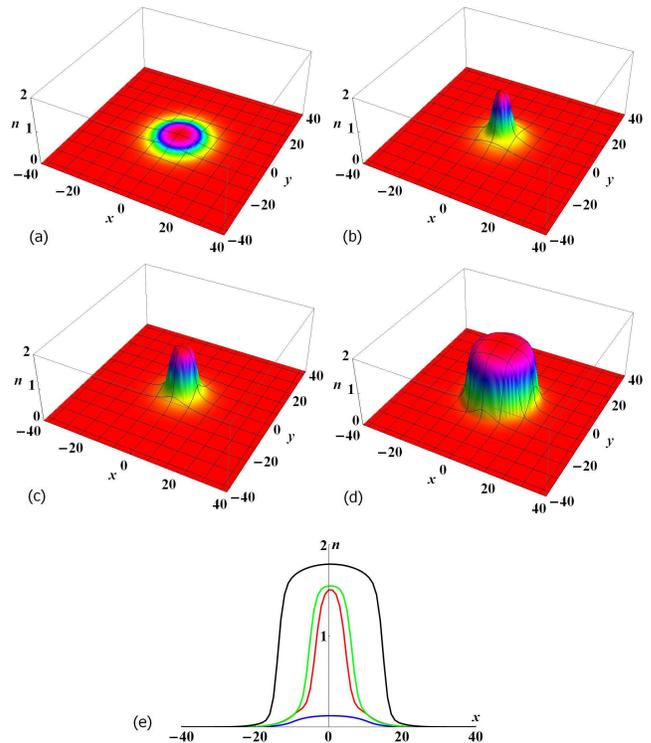}}\caption{(color online). The distribution of the indirect excitons density in the quantum well plane in case of the pumping (\ref{eq6}) with changing the pumping intensity in the case of narrow laser beams, $G_0$: a) 0.0015; b) 0.0055; c) 0.0075; d) 0.035. The shape parameter is $\sigma=10$. The parameters of the system (in
dimensionless units) are $d_1$=0.1, $b_1$=-2.3, $\tau$=100. e) The density profile of the pumping through the center of the system for the set of pumping `a-d'.}\label{fig1}\end{figure}

Fig.~1 is plotted in dimensionless variables. If one selects $\tau$=10 ns, $T_{\rm 2}$=6 K, $T$=2K,
$D=4\cdot10$ cm$^2$/s, $l_{\rm u}$=0.63 $\mu$m, $n_{\rm u}$=$3.2\cdot10^{10}$ cm$^{-2}$,
$an_{\rm u}$=1.725 meV, $\sigma$=6.3 $\mu$m, then a peak value of the exciton density corresponding the condensed phase is $n$=$2.1\cdot10^{10}$ cm$^{-2}$.

Now let analyze the structure formation of the exciton emission in case of a wider beam, sizes of which are comparable with the location of several islands of the condensed phase. The calculatiuons with increasing the pumping intensity, presented in Fig.~2, reveal changes in  the exciton density distribution from the gas phase (Fig.~2a) to more complicated structures: the appearance of the islands of the condensed phase (Fig.~2b) and the transformation of the islands into a continuous emission ring  (Fig.~2c).

\begin{figure}\centerline{\includegraphics[width=8.6cm]{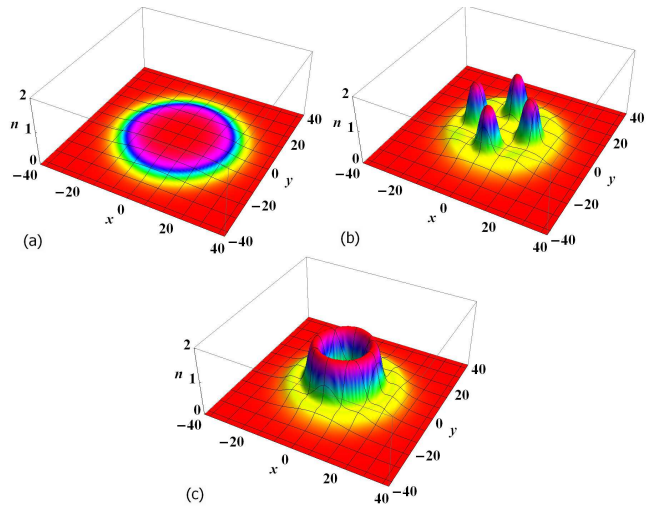}}\caption{(color online). The distribution of the exciton density in case of the pumping (\ref{eq6}) with changing the pumping intensity $G_0$: a) $ 0.0015; b) $ 0.0045; c) 0.0065. The shape parameter is $\sigma$=24. The parameters of the system are the same as in Fig.~1.}\label{fig2}\end{figure}

If the numerical values of the parameters are chosen similar to the given in the previous task, the size of the laser spot will be 15.2 $\mu$m, the peak value of the pumping will remain the same.

\section{Dependence of the exciton density structure on laser beam width at fixed pumping density $G_0$}Separately, let analyze the modification of the structure of the exciton density with increasing the shape parameter $\sigma$ at a fixed value of  $G_0$. In this case the beam becomes wider, the intensity of pumping in the center of the beam persists the same, but the whole pumping increases. The value of $G_0$  is chosen in the interval, in which the transition from the gas phase to the condensed one occurs and where the formation of the excitons islands is possible. The calculations, presented in Fig.~3, show that the number of formed islands increases in case of wider pumping spots. The several island rings appear at increasing the laser spot. Multiple rings are observed in the experiment~\cite{1}.

\begin{figure}[ph]\centerline{\includegraphics[width=8.6cm]{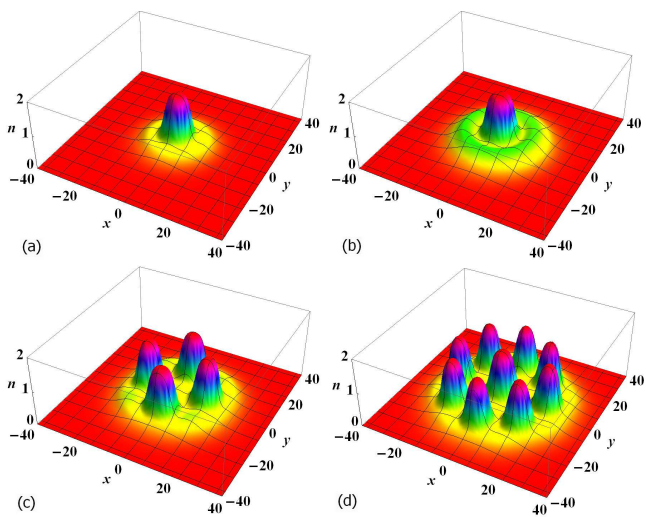}}\caption{(color online). The distribution of the exciton density with increasing the width of the laser beam $\sigma$: a) 14; b) 20; c) 24; d) 30. $G_0$=0.0055 is fixed for all images. The parameters of the system are the same as in Fig.~1}\label{fig3}\end{figure}

We calculated the dependence of the distance from the islands to the center of the structure on the exciton lifetime and on the pumping. The exciton lifetime may be increased by increasing the applied voltage~\cite{1}. According to the calculations, the distance from the islands to the  center  of the structure (Fig.~2b) shifts from 13 to 14 with changing the pumping from 0.0045 to 0.0055 (all units are dimensionless). The Fig.~3c illustrates a similar shift of the  islands-center distance from 13 to 14 with changing  the lifetime from 100 to 130. The results showed that  in  both cases islands move away from the center in accordance with the experimental results of the work~\cite{1}, where a slow increase of the inner ring radius was observed. Further increasing the pumping or the exciton lifetime may cause a restructuring of the  pattern:  appearance of  continuous rings, a second ring of the islands and so on (see Figs.~2 and 3).
\section{Structures emerging in case of irradiation by two lasers}Another phenomenon observed in Ref.~\cite{1} should be considered. Let analyze the behavior of the system in case of the irradiation by two laser beams, the shape of each is given by the Equation (\ref{eq6}). The beams are spatially separated in the plane of the quantum wells. Assume that the radiation power is such that the structure arises in the form of a continuous ring or in the form of islands of the condensed phases of excitons under the action of each beam. We calculated the change in the structure with shortening the distance between the centers of the laser spots. Such dynamics of these changes were measured in Ref.~\cite{1}.

The results of the solution of Equation (\ref{eq5}) with the pumping, created by two lasers, are shown in Fig.~4. As one can see, there are two independent structures of the condensed phases of islands, created by individual lasers, for large distances between the centers of the laser spots. As the distance between the centers decreases (the left spot approaches to the right one), the structure of islands location is slightly deformed, and at the coincidence of the centers, the structure with a common center arises. Such a pattern was observed in the experiment~\cite{1}. This demonstrates the plausibility of the model.

\begin{figure}[pt]\centerline{\includegraphics[width=8.6cm]{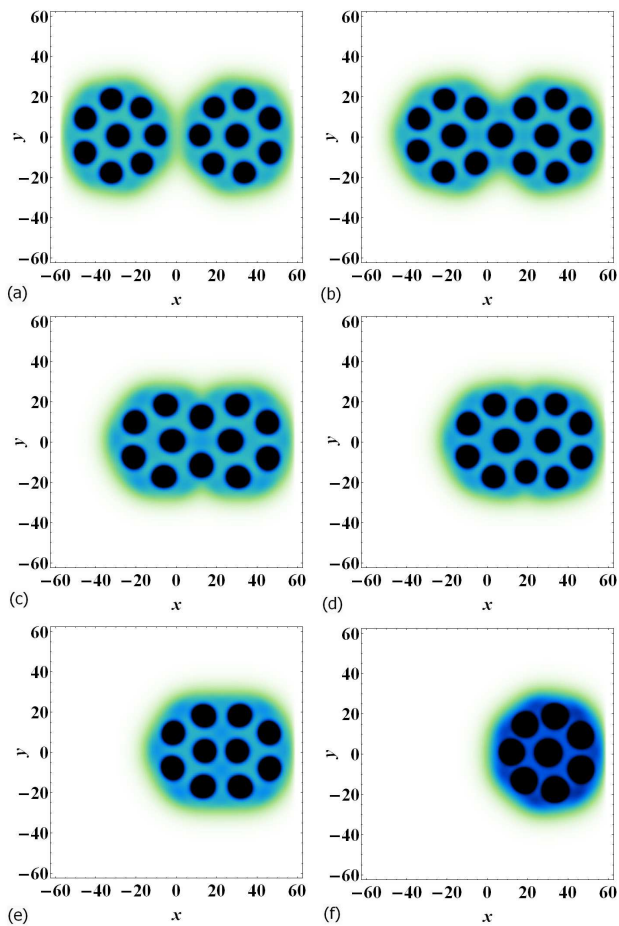}}\caption{(color online). The distribution of the exciton density in the plane of the quantum well in case of the pumping by two spatially separated
sources, the distance $r_{12}$ between the centers of which is: a) 60.5; b) 48.5; c) 36.5; d) 5.24 ; e) 16.5; f) 0. The patrameters of the system are $G_0$=0.0055, $d_1$=0.1, $\tau$=100, $b_1$=-2.3, $T$=2 K, $\sigma$=28.}\label{fig5}\end{figure}

Fig.~4 is plotted in dimensionless variables. If we choose $\tau$=10 ns, $T_{\rm 2}$=6 K, $T$=2 K, $D$=4 cm$^2$/s, $l_{\rm u}$= 0.63$\mu$m, $n_{\rm u}$=$3.2\cdot10^{10}$cm$^{-2}$, $an_{\rm u}$=1.725 meV, then $\sigma$=17.7 $\mu$m and the distance between the centers of the pumping sources in the figure is, $\mu$m: a) 38.6 ; b) 30.7; c) 23.1; d) 15.5; e) 10.4; e) 0.

There is a difference between the manifestations of the merging of two external rings and two inner rings. The region without islands appears  between the laser spots in the case of two external rings at close distances (see the experiment~\cite{2} and the calculation of Ref.~\cite{Pol}). The following arguments give a qualitative explanation of this fact. The excitons of the external ring are formed when the electrons provided by the crystal donors bind with holes arriving from the laser sport. In the region between the two external rings, the number of the donor electrons is not sufficient for the creation of the islands for both rings, thus the region depleted of islands emerges. In the case of the inner ring, excitons  arise directly from the electrons and holes created by laser and such region depleted of islands does not appear. Such qualitative explanation is confirmed by the calculations for the inner rings (see Fig.~4 in this work) and for the external rings (see Fig.~3 in Ref.~\cite{Pol}). Both these calculations are carried out using the same model as presented above.
\section{Role of the temperature}Let investigate the dependence on the temperature in a phenomenological model. According to the Landau model, the coefficient $b_1$ in the free energy (\ref{eq3}) can be expanded  in the vicinity of the point $T_2$, at which the minimum of the free energy disappears:
\begin{equation}\label{eq20}b_1=b_{\rm 1,c} +B\left( {1 - \frac{T}{T_{\rm 2}}} \right)T_{\rm 2},\end{equation}
\N where $b_{\rm 1,c}=b_1\left(T_{\rm 2}\right)=-2$.

The solution of the Equation (\ref {eq5}) using Eq.~(\ref{eq20}) shows that with increasing the temperature the individual islands of the condensed phase become narrower (Fig.~5b,c) and at a certain temperature the condensed phase disappears (Fig.~5d).
\begin{figure}[pt]\centerline{\includegraphics[width=8.6cm]{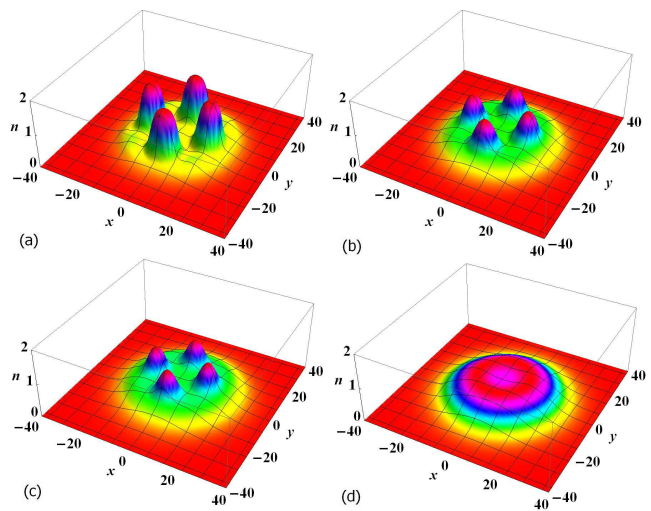}}\caption{(color online). The density distribution of indirect excitons at the same pumping $G_0$=0.0055 and the pumping width $\sigma$=24 with increasing the temperature $T$, K: 1) 2; 2) 3; 3) 3.5; 4) 4; 5) 5; 6) 7. The critical temperature $T_{\rm 2}$=6 K, the other parameters of the system are the same as in Fig.~3.}\label{fig4}\end{figure}

The temperature is determined by the thermostat and caused by the laser heating. Moreover, the temperature of the crystal lattice may diverge from the exciton subsystem temperature \cite{Ivan}.

The calculations of the lattice heating taking into account the thermal conductivity of GaAs~\cite{36} gave a small temperature deviation from the bath temperature at radiation intensities and the spot sizes, used in our calculations presented above. The calculations of the exciton temperature based on the exciton relaxation processes by phonons \cite{Ivan} showed that the temperature of the exciton subsystem may be different from the lattice temperature by a few degrees and is inhomogeneous within the spot. In order to decrease the effects of crystal heating, the calculations were performed for not great sizes of the laser spots and for small laser power values (but the values exceeded the threshold of the condensed phase arising). Herewith, the structure is formed at the border of the region, where the temperature inhomogeneity is not significant.

\section{Conclusions}This paper gives an explanation to the phenomenon found in Ref.~\cite{1} of the fragmentation of the exciton emission from the quantum wells under the laser irradiation. The explanation is based on a model basing on the following assumptions: 1) the existence of the exciton condensed phase, caused by the attraction between excitons; 2) the essential role of the finite lifetime of excitons.

The periodic distribution of the exciton density along the inner ring occurs if the region of the laser spot is bigger than a certain value and the intensity of the laser exceeds a certain threshold value. The structure formation is a process of self-organization in the non-equilibrium system. With increasing the temperature, the structure becomes less distinct and, finally, disappears. When two inner rings, created by two lasers, are converged, there is a mutual deformation of the rings and there is the coherence in the density distribution.

Thus, the experiments with indirect excitons are explained in the present paper basing the folllowing model: besides the long-range repulsive interaction between the excitons, there is a region of attractive interaction at smaller distances. As it was pointed out in Sect.~I, this model allowed to explain a series of other various phenomena, related with indirect excitons in quantum wells. Large variety of experimental evidence, which are confirmed by the considered model, demonstrates its validity. Note that the exciton system with its fast kinetics and easiness of creation can be the base model for the simulation of the phase transitions in the different systems of unstable particles -- for an electron-hole liquid, structures formation in radiation physics, a gas with a large number of excited molecules, for the quark-gluon plasma etc.

\bibliography{apssamp}\end{document}